\global\def\draftcontrol{0}

   \def\versionno{From Courant to actions}
\catcode`\@=11

\expandafter\ifx\csname draftcontrol\endcsname\relax\global\def\draftcontrol{0}
\fi

{\count255=\time\divide\count255 by 60
\xdef\hourmin{\number\count255}
\multiply\count255 by-60\advance\count255 by\time
\xdef\hourmin{\hourmin:\ifnum\count255<10 0\fi\the\count255}}
\def\draftdate{\number\month/\number\day/\number\year\ \ \ \hourmin }

\newcommand\makepapertitle{\par
  \begingroup
    \renewcommand\thefootnote{\@fnsymbol\c@footnote}%
    \def\@makefnmark{\rlap{\@textsuperscript{\normalfont\@thefnmark}}}%
    \long\def\@makefntext##1{\parindent 1em\noindent
            \hb@xt@1.8em{%
                \hss\@textsuperscript{\normalfont\@thefnmark}}##1}%
     \newpage
     \global\@topnum\z@   % Prevents figures from going at top of page.
     \@makepapertitle
     \thispagestyle{empty}\@thanks
  \endgroup
  \setcounter{footnote}{0}%
  \global\let\thanks\relax
  \global\let\makepapertitle\relax
  \global\let\@makepapertitle\relax
  \global\let\@thanks\@empty
  \global\let\@author\@empty
  \global\let\@date\@empty
  \global\let\@title\@empty
  \global\let\title\relax
  \global\let\author\relax
  \global\let\date\relax
  \global\let\and\relax
  \def\version{\let\version\@version\@gobble}
}
\def\@makepapertitle{%
  \newpage
   \ifnum\draftcontrol=1 {}
   \version\versionno
   \vskip 3em%
   \else
   \hfill\hbox to 3cm {\parbox{4cm}{\@pubnum}\hss}%
   \vskip 3em%
   \fi
   \begin{center}%
   \let \footnote \thanks
     {\LARGE {\@title}}%
     \vskip 1.5em%
     {\normalsize%\large
       \lineskip .5em%
       \begin{tabular}[t]{c}%
         \@author
       \end{tabular}\par}%
     \vskip 1.5em%
     {\@bstract}%
     \end{center}%
     \vskip 1.5em
     \@date%
   \par
}

\gdef\@pubnum{}
\def\pubnum#1{%
  \gdef\@pubnum{#1}}

\gdef\@bstract{}
\def\Abstract#1{%
  \gdef\@bstract{%
   \parbox{\textwidth-0pc}{%
   \centerline{\bf Abstract}\penalty1000%
\noindent%\abstractfont \baselineskip=12pt
\renewcommand\baselinestretch{1.0}%
{#1}}}
}

\def\ps@paper{\let\@mkboth\@gobbletwo%
     \ifnum\draftcontrol=1
        \def\@oddfoot{\hbox to \textwidth{\tiny \versionno \hfil\tiny\draftdate}%
        \hskip -\textwidth \hbox to \textwidth{\hfil\rm\thepage\hfil}}%
     \else\def\@oddfoot{\hbox to \textwidth{\hfil\rm\thepage\hfil}}
     \fi
     \let\@evenfoot\@oddfoot
}
\def\@version#1{\ifnum\draftcontrol=1
\typeout{}\typeout{#1}\typeout{}
\vskip3mm\centerline{\hbox{\fbox{\normalsize{\tt DRAFT -- #1 -- }
                   {\draftdate}}}}\vskip3mm
\fi}
\let\version\@version
\long\def\eqlabel#1{\ifnum\draftcontrol=1
                    \tag@false  % there are some problems with multline without this
                    \tag*{(\theequation) \hbox to -0.2cm{\hspace{0cm}\small{#1}\hss}}
                    \refstepcounter{equation}
                    \edef\@currentlabel{\theequation}
                    \ltx@label{#1}          % use old LaTeX \label instead of new definition
                    \else
                    \label{#1}
                    \fi
                    }
\let\st@bibitem\@bibitem
\let\st@lbibitem\@lbibitem
\ifnum\draftcontrol=1
  \def\@bibitem#1{%
    \st@bibitem{#1}\a@@label{#1}\ignorespaces}
  \def\@lbibitem[#1]#2{%
    \st@lbibitem[#1]{#2}\a@@label{#2}\ignorespaces}
  \def\a@@label#1{%
    \gdef\a@lab{\smash{\normalfont\small#1}}
    \ifvmode
      \if@inlabel
        \global\setbox\@labels\hbox{%
          \llap{\a@lab\let\a@lab\relax
                \kern\@totalleftmargin\kern\marginparsep}%
          \box\@labels}%
      \fi
    \fi}
\fi
\documentclass[12pt,letterpaper]{article}
\usepackage{amsmath,amssymb,array,calc,rotating,epsfig,psfrag}
\usepackage[nosort]{cite}
\usepackage{graphicx}
\usepackage{hyperref}
\usepackage{mathtools,slashed}

\ifnum\draftcontrol=1
\fi
\renewcommand\baselinestretch{1.25}
\renewcommand\section{\@startsection {section}{1}{\z@}%
                                   {-3.5ex \@plus -1ex \@minus -.2ex}%
                                   {2.3ex \@plus.2ex}%
                                   {\normalfont\large\bfseries}}
\renewcommand\subsection{\@startsection{subsection}{2}{\z@}%
                                   {-3.25ex\@plus -1ex \@minus -.2ex}%
                                   {1.5ex \@plus .2ex}%
                                   {\normalfont\normalsize\bfseries}}
\renewcommand\subsubsection{\@startsection{subsubsection}{3}{\z@}%
                                   {-3.25ex\@plus -1ex \@minus -.2ex}%
                                   {1.5ex \@plus .2ex}%
                                   {\normalfont\normalsize\it}}
\renewcommand\paragraph{\@startsection{paragraph}{4}{\z@}%
                                   {-3.25ex\@plus -1ex \@minus -.2ex}%
                                   {1.5ex \@plus .2ex}%
                                   {\normalfont\normalsize\bf}}

\def\revise#1       {\raisebox{-0em}{\rule{3pt}{1em}}%
                     \marginpar{\raisebox{.5em}{\vrule width3pt\
                     \vrule width0pt height 0pt depth0.5em
                     \hbox to 0cm{\hspace{0cm}{%
                     \parbox[t]{4em}{\raggedright\footnotesize{#1}}}\hss}}}}

\catcode`\@=12

\usepackage[active]{srcltx}
\usepackage{color}

\begin{document}

%%%%%%%%%%%%%%%%%%%%%%%%%%%%%%%%%%%%%%End of Alex Draft Mode%%%%%%%%%%%%%%%%%%%%

%%%%%%%%%%%%%%%%%%From Carlo's Style%%%%%%%%%%%%%%%%%%%%%%%%%%%%%%%%%%5

%\overfullrule=0pt
%\parskip=2pt
%\parindent=12pt
%\headheight=0in
%\headsep=0in
%\topmargin=0.50in
%\oddsidemargin=0in

%--------+---------+---------+---------+---------+---------+---------+
\newcommand{\al}[1]{\begin{align}#1\end{align}}
\newcommand{\all}[1]{\begin{align*}#1\end{align*}}
\newcommand{\be}{\begin{equation}}
\newcommand{\ee}{\end{equation}}
\newcommand{\beq}{\begin{equation}}
\newcommand{\eeq}{\end{equation}}
\newcommand{\ba}{\begin{eqnarray}}
\newcommand{\ea}{\end{eqnarray}}
\newcommand{\nn}{\nonumber}
%--------+---------+---------+---------+---------+---------+---------+

% Ofer's definitions
\def\vol{\bf vol}
\def\Vol{\bf Vol}
\def\del{{\partial}}
\def\vev#1{\left\langle #1 \right\rangle}
\def\cn{{\cal N}}
\def\co{{\cal O}}
%\newfont{\Bbb}{msbm10 scaled 1200}     %instead of eusb10
%\newcommand{\mathbb}[1]{\mbox{\Bbb #1}}
\def\IC{{\mathbb C}}
\def\IR{{\mathbb R}}
\def\IZ{{\mathbb Z}}
\def\RP{{\bf RP}}
\def\CP{{\bf CP}}
\def\Poincare{{Poincar\'e }}
\def\tr{{\rm tr}}
\def\tp{{\tilde \Phi}}
\def\Y{{\bf Y}}
\def\te{\theta}
\def\bX{\bf{X}}

\def\TL{\hfil$\displaystyle{##}$}
\def\TR{$\displaystyle{{}##}$\hfil}
\def\TC{\hfil$\displaystyle{##}$\hfil}
\def\TT{\hbox{##}}
\def\HLINE{\noalign{\vskip1\jot}\hline\noalign{\vskip1\jot}} %Only in
\def\seqalign#1#2{\vcenter{\openup1\jot
  \halign{\strut #1\cr #2 \cr}}}
\def\lbldef#1#2{\expandafter\gdef\csname #1\endcsname {#2}}
\def\eqn#1#2{\lbldef{#1}{(\ref{#1})}%
\begin{equation} #2 \label{#1} \end{equation}}
\def\eqalign#1{\vcenter{\openup1\jot   }}
\def\eno#1{(\ref{#1})}
\def\href#1#2{#2}
\def\half{{1 \over 2}}

%--------+---------+---------+---------+---------+---------+---------+
%Hirosi's macros:
\def\ads{{\it AdS}}
\def\adsp{{\it AdS}$_{p+2}$}
\def\cft{{\it CFT}}

\newcommand{\ber}{\begin{eqnarray}}
\newcommand{\eer}{\end{eqnarray}}

\newcommand{\bea}{\begin{eqnarray}}
\newcommand{\eea}{\end{eqnarray}}

\newcommand{\beqar}{\begin{eqnarray}}
\newcommand{\cN}{{\cal N}}
\newcommand{\cO}{{\cal O}}
\newcommand{\cA}{{\cal A}}
\newcommand{\cT}{{\cal T}}
\newcommand{\cF}{{\cal F}}
\newcommand{\cC}{{\cal C}}
\newcommand{\cR}{{\cal R}}
\newcommand{\cW}{{\cal W}}
\newcommand{\eeqar}{\end{eqnarray}}
\newcommand{\lm}{\lambda}\newcommand{\Lm}{\Lambda}
\newcommand{\eps}{\epsilon}

%--------+---------+---------+---------+---------+---------+---------+

\newcommand{\nonu}{\nonumber}
\newcommand{\oh}{\displaystyle{\frac{1}{2}}}
\newcommand{\dsl}
  {\kern.06em\hbox{\raise.15ex\hbox{$/$}\kern-.56em\hbox{$\partial$}}}
\newcommand{\as}{\not\!\! A}
\newcommand{\ps}{\not\! p}
\newcommand{\ks}{\not\! k}
\newcommand{\D}{{\cal{D}}}
\newcommand{\dv}{d^2x}
\newcommand{\Z}{{\cal Z}}
\newcommand{\N}{{\cal N}}
\newcommand{\Dsl}{\not\!\! D}
\newcommand{\Bsl}{\not\!\! B}
\newcommand{\Psl}{\not\!\! P}
\newcommand{\eeqarr}{\end{eqnarray}}
\newcommand{\ZZ}{{\rm \kern 0.275em Z \kern -0.92em Z}\;}

%%%%%%%%%%%%%%%%%Some more definitions%%%%%%%%%%%%%%%%%%%5
\def\s{\sigma}
\def\a{\alpha}
\def\b{\beta}
\def\r{\backslash l}
\def\d{\delta}
\def\g{\gamma}
\def\G{\Gamma}
\def\ep{\epsilon}
%%%%% number equations by section %%%%%%%%
\makeatletter \@addtoreset{equation}{section} \makeatother
\renewcommand{\theequation}{\thesection.\arabic{equation}}
%%%%%%%%%%%%%%%%%%%%%%%%%%%%%%%%%%%%%%%

%%%%%%%%%%%%%%Leo's%%%%%%%%%%%%%%%%%%%%5
\def\be{\begin{equation}}
\def\ee{\end{equation}}
\def\bea{\begin{eqnarray}}
\def\eea{\end{eqnarray}}
\def\m{\mu}
\def\n{\nu}
\def\g{\gamma}
\def\p{\phi}
\def\L{\Lambda}
\def \W{{\cal W}}
\def\bn{\bar{\nu}}
\def\bm{\bar{\mu}}
\def\bw{\bar{w}}
\def\ba{\bar{\alpha}}
\def\bb{\bar{\beta}}

\begin{titlepage}

\vskip 1.7 cm

\centerline{\bf \Large  The Diffeomorphism Field Revisited }

\vskip 1.5cm 
\centerline{\large Delalcan Kilic }

\vskip 1cm
\centerline{\it Department of Physics and Astronomy}
\centerline{ \it  The University of Iowa}
\centerline{\it Iowa City, IA 52242}
\centerline{\it delalcan-kilic@uiowa.edu}

\vspace{1cm}

\begin{abstract}

We revisit the dynamical action, developed in earlier studies \cite{BRY,RY,Takthesis}, of the  gravitational analog of Yang-Mills  field, called the diffeomorphism field. 
We show an inconsistency in the construction of this action and solve it by a modification. The modified action becomes structurally similar to the Yang-Mills action.  Then we explain the problems that arise from interpreting the diffemorphism field as a tensor. We leave out this interpretation, thereby also covariantization as a method to reach diffeomorphism invariance.  By introducing corrections that involve affine connection coefficients we recover full diffeomorphism invariance in interactions of the diffeomorphism field with other fields.  We also show that this approach maintains the spatial diffeomorphism invariance of the theory itself.   
\end{abstract}
\end{titlepage}
 %%%%%%%%%%%%%%%%%%%%%%%%%%%%%%%%%%%%%%%%%%%%%%%%%%%%%%%%%%%%%%%%%%%%%%%%%%

\section{Introduction}
Wess-Zumino-Witten (WZW) action\cite{WZNW}  and Polyakov's two-dimensional quantum gravity (P2DG) action in lightcone gauge \cite{Pol87} can be obtained as parts of the geometric actions built on the coadjoint orbits of Kac-Moody (KM) and Virasoro algebras, respectively\cite{RR89,AS89}.  Additional terms appear on each sector. On the KM sector the additional term is equivalent\cite{diVecetal} to the coupling of a  background Yang-Mills (YM) field to the WZW field. In this equivalence  the KM coadjoint element $A$ is identified with the space component $A_1$ of the YM field $A_{\mu}$, in the temporal gauge $A_0=0$. On the Virasoro sector the additional term is equivalent to the coupling of a background field $D$ to the Polyakov field.
The field $D$  can be seen as the space-space component $D_{11}$ of a rank two field  $D_{\mu \nu}$ (whose coordinate transformation deviates from that of a  tensor)  in the temporal gauge $D_{01}=0$. The $D_{00}$ component is simply invisible due to the lightcone gauge of the metric used in P2DG action, and  is at our disposal. The field $D_{\mu \nu}$ is  used to be called the diffeomorphism field (in short, the diff field).
 
Classical gravity is dynamically trivial in 2D. Dynamics of the spacetime metric comes from anomaly contribution to the quantum effective actions of matter fields coupled to gravity.  P2DG action is originally obtained as the effective action with conformal anomaly\cite{Pol81}.   P2DG action in lightcone gauge is regarded as the gravitational analog of the WZW action\cite{Zamo}.  The diff field is, in the same sense, the gravitational analog of a YM field in 2D. If this holds to be true in higher dimensions, then the diff field  may have an important role in gravity in higher dimensions. In particular, it may provide an alternative description for the graviton. With a similar motivation in mind,  a dynamical theory for the diff field is constructed in Refs\footnote{One can also check \cite{delotez} for an extensive review of all.} \cite{LR95,BLR,BRY,RY,Takthesis}.  These studies mainly focus on examinining the relationship between the KM coadjoint representation and the YM theory in 2D, and mimic it to build the analogous action for the Virasoro coadjoint representation.

 In this paper we focus on the latest approach\cite{BRY,RY,Takthesis} used to construct the so called transverse actions associated with KM and Virasoro coadjoint representation.  We review this construction in Section \ref{TransverseMain}. In the stated references  the YM form of the momentum was directly assumed to recover the YM action as the KM transverse action. Similarly, the conjugate momentum to diff field was chosen to have a specific form to build the diff field action. Here we show that the general form of the momenta used to build the actions is dictated by the Gauss law constraint (and its diff field analog) for the construction to be self-consistent. In fact, the momentum obtained from the constructed diff field action in Refs \cite{BRY,RY,Takthesis} is not the same as the momentum chosen to construct the action. We introduce a simple ansatz for the momentum  which yield the YM form of the momentum automatically from the constructed KM transverse action. Also, the analogous ansatz for the diff field case solves the stated consistency problem.

Another problem that we observe in the previous studies \cite{BRY,RY,Takthesis} is that the diff field is assumed to transform as a tensor, so in order to reach a spacetime diffeomorphism invariant one simply covariantizes  the constructed flat-space action. We explain why this approach is problematic in Section \ref{Covariantization}. We abandon the idea of a tensorial diff field. The constructed flat-space action is shown to be  spatial-diffeomorphism invariant provided the diff field does not transform as a tensor.  We introduce a way to obtain a spacetime tensor from the diff field in Sections \ref{connectionVircoadjointelement},\ref{spatialtensor2D} and \ref{DiffInvRecovered} . By this method 
we recover spacetime diffeomorphism invariance of the diff field in its interactions with other fields (introduced in \cite{BRY}). This requires a change in the higher dimensional lift of the Virasoro coadjoint action. Also, the diff momentum picks up corrections involving affine connection coefficients as does the action. Nevertheless, the structural similarity with the YM action and the spatial diffeomorphism invariance of the action are preserved. We do not provide a fully diffeomorphism invariant action for the diffeomorphism field in this work. We believe this requires the introduction of fields analogous to the lapse and shift functions of the ADM formalism of general relativity.

\section{Coadjoint Actions }	
The coadjoint actions of the semidirect product of KM and Virasoro algebras are given by\cite{LR95}
\al{ \label{KMcoad}
\delta A & = \xi A' + \xi' A - [\Lambda , A] + k^{-1} \Lambda', \\
\delta D & = \xi D' + 2 \xi' D + q \xi''' - \text{Tr}(A \Lambda') \label{Vircoad}
}
where prime denotes derivative with respect to the coordinate $x^1$ of the circle over which the algebras are defined. $\Lambda$ and $\xi$ are adjoint elements,  and $A$ and $D$ are coadjoint elements of KM and Virasoro algebras, respectively. Trace is over the lie algebra of the base group of the KM algebra, and $k$ and $q$ are constants related to the centers of the algebras.

 The first two terms  of \eqref{KMcoad} represent the infinitesimal coordinate transformation (or the lie derivative) of a YM field $A$ in 1D. Alternatively, it can be seen as the infinitesimal, spatial and time-independent coordinate transformation  of the spatial  component $A_1$ of a YM field $A_{\mu}$ in 2D, that is, under $x^{\mu} \rightarrow x^{\mu} - \xi^{\mu}(x)$ with 
 \al{ \label{spatialtimeindepxi}
 \xi^0 = 0 = \partial_0 \xi^1.
 }

 Similarly, the first two terms of \eqref{Vircoad} can be considered as the infinitesimal coordinate transformation of a rank-two tensor in 1D. Alternatively, it can be regarded as the infinitesimal, spatial, time-independent coordinate transformation, \eqref{spatialtimeindepxi}, of the space-space component $D_{11}$ of a rank-two tensor $D_{\mu \nu}$ in 2D. Notice, however, that the third term in \eqref{Vircoad} also involves $\xi$, so it must be part of the coordinate transformation. Hence, $D_{\mu \nu}$ does not transform as a rank-two tensor. 

Next, we focus on the terms involving the KM adjoint element $\Lambda$. The last two terms  of \eqref{KMcoad} can be seen as the infinitesimal gauge transformation of a YM field in 1D. That is,  we consider $A \rightarrow g A g^{-1} + k^{-1} g d g^{-1}$ and $g  \approx I - \Lambda \equiv I + i T^a \Lambda^a$.  Alternatively, it can be seen as the time-independent gauge transformation ($\partial_0 \Lambda=0$) of $A_1$ component of a YM field in 2D. 

One can also observe that there is a KM contribution in \eqref{Vircoad} (the trace term). This shows that the object $D$ is not gauge-invariant in the presence of a YM field. It is straightforward\cite{LR95} to remedy this, namely, one can define 
\al{ \label{gaugeinvdiff}
\tilde{D} = D + (k/2) \text{Tr}(A A).
}
This field is gauge invariant, $\delta_{\Lambda} \tilde{D} = 0$.

\section{Construction of Dynamical Actions for the Coadjoint Elements} \label{TransverseMain}
To construct dynamical actions for the coadjoint elements $A$ and $D$ one uses the isotropy equations, which are obtained by setting the coadjoint actions \eqref{KMcoad}, \eqref{Vircoad} to zero, i.e. $\delta A = 0 = \delta D$. These equations define the adjoint elements $\xi, \Lambda$ which fix  the given coadjoint elements $D, A$ on their orbits. If we imagine infinitely many identical copies of  coadjoint orbits on top of each other forming constant time slices of a 2D spacetime, then the isotropy equation would define the directions that are transverse to the slices. This is the motivation of the term transverse action.

\subsection{From Kac-Moody Algebra to Yang-Mills Action}
If $A$ is to be identified with a YM field, we should recover the YM action from the construction. 
The primary observation in\cite{LR95} was that the pure KM part of the KM isotropy equation,
\al{ \label{pureKMcoadKM}
 0=\delta_{\Lambda} A =k^{-1}  \Lambda' - [\Lambda , A],
 }
lifts to the Gauss law constraint $G$ under $\Lambda \rightarrow k E$, where $E$ is the conjugate momentum to $A$. To make this formal, one looks\cite{RY} for a charge $Q$ that generates the transformation  i.e.
\al{ \label{KMGaussgenerator}
\delta_{\Lambda} A = \{ A , Q \}
} 
where $\{ \ , \}$ is the Poisson bracket (PB) for the corresponding field theory and the charge is explicitly given by $Q = - k^{-1} \int dx^1  \Lambda \, G$. Using \eqref{pureKMcoadKM}   one recovers the Gauss law operator $G$ from \eqref{KMGaussgenerator}
\al{
G  =  E' + k [A , E] = \partial_1 E^1 +  k[A_1 , E^1].
}
Notice that the momentum $E^1$ has a hidden time index coming from its definition, 
\al{ \label{E1fromS}
E^1 \equiv \delta S / \delta (\partial_0 A_1).
}
Thus, the operator $G$  also contains a time index. We write $E^{10}$ and $G^0$ from here on. 

The prescription to obtain the dynamical action \cite{Takthesis} is given by
\al{ \label{transactpresc}
\mathcal{L} = \text{ST} - \mathcal{H} + \lambda_0 G^0.
}
Here ST stands for the symplectic term $E^{10} \partial_0 A_1$, $\mathcal{H}$ is the Hamiltonian density $E_{10} E^{10}/2$, $\lambda_0$ is the Lagrange multiplier for the Gauss law constraint, and the trace over the algebra is suppressed. We pick $\lambda_0=cA_0$ as this is the simplest choice with a single  time index. 
The prescription \eqref{transactpresc} is simply that of a constrained field theory. The Gauss law enforces the isotropy equation i.e. the dynamics is enforced to be transverse to the orbits, not within one.   The only missing piece to proceed is the expression for the momentum $E$. 
  
  In \cite{RY,Takthesis}  the YM form of the momentum $E$ is directly assumed to recover the YM action as the KM transverse action, not deduced from the method itself. We do not want to do this since  we do not have an expression ready at hand in the Virasoro case. Hence, at this point we introduce a fairly relaxed ansatz for the momentum, 
\al{ \label{momansKM}
E^{10} = -\partial^0 A^1 + M^{10}
}
where $M^{10}$ is assumed to be independent of the time derivative of $A_{\mu}$. We put  the components into \eqref{transactpresc} and obtain $\mathcal{L}$ in terms of  $A_0, A_1, M_{10}$. Then we recompute the momentum $E$ from the constructed action $S = \int d^2 x \, \mathcal{L}$ using \eqref{E1fromS}. The result is 
\al{
E^{10} = - \partial^0 A^1 + c \partial^1 A^0 - c k [A^0, A^1].
} 
This is the  momentum $F_{01} \equiv E_{01}$ conjugate to $A_1$  for $c=1$. Setting $c$ to any real constant is legitimate as it amounts to rescaling $A_0$,  which is nondynamical. The meaning of setting $c=1$, on the other hand, is that the Gauss law is the constraint associated\footnote{The Gauss law is the field equation of $A_0$. Equivalently, it is the secondary constraint that follows from the vanishing of the conjugate momentum of $A_0$.} with $A_0$, not with an arbitrary multiple of it. Notice also that $M^{10}$ does not depend on velocities $\partial^0 A^{\mu}$ so the ansatz \eqref{momansKM} is consistent. 

If we insert the explicitly found momentum back into the Lagrangian we get  the YM Lagrangian in 2D up to a spatial boundary term $\partial_1 (A_{0a} F^{10a})$. We assume the necessary boundary conditions to make this term vanish or consider a spacetime whose spatial hypersurface is closed. One can straightforwardly extend the Lagrangian to higher dimensions and covariantize.

\subsection{From Virasoro Algebra to the Diff Field Action} \label{Virasorotransverse}
The pure Virasoro part of the coadjoint action \eqref{Vircoad} reads 
\al{ \label{deltaxiD}
\delta_{\xi} D = \xi D' + 2 \xi' D + q \xi'''
}
and the isotropy equation is $0=\delta_{\xi} D$. 
We lift this transformation to $N+1$ dimensions. The most straightforward lift 
reads\cite{Takthesis}
\al{ \label{CoadNDtrf}
\delta_{\xi} D_{\mu \nu} = \xi^{\lambda} \partial_{\lambda} D_{\mu \nu} + \partial_{\mu} \xi^{\lambda} D_{\lambda \nu} + \partial_{\nu} \xi^{\lambda} D_{\mu \lambda} + q \partial_{\mu} \partial_{\nu} \partial_{\lambda} \xi^{\lambda}.
}
For convenience we assume that the diff field $D_{\mu \nu}$ to be symmetric (only the symmetric part admits the term with coefficient $q$).  
For a spatial, time-independent coordinate transformation  \eqref{spatialtimeindepxi}
this yields
\begin{subequations}\label{allDscoadhigher}
\al{ \label{higherdimcoad}
\delta_{\xi} D_{ij} & = \xi^k \partial_k D_{ij} + \partial_i \xi^k D_{kj} + \partial_j \xi^k D_{ik}+ q \partial_i \partial_j \partial_k \xi^k, \\
\delta_{\xi} D_{0i}  & = \xi^k \partial_k D_{0i} + D_{0k} \partial_i \xi^k, \\
\delta_{\xi} D_{00}  & = \xi^k \partial_k D_{00}
}
\end{subequations}
with $i,j,k = 1, \cdots, N$.  In 2D, i.e. when $N=1$, \eqref{higherdimcoad} is equivalent to \eqref{deltaxiD} with $D \equiv D_{11}$ and  $\xi \equiv \xi^1$. Note that these transformations preserve the full temporal gauge $D_{0k}=0 = D_{00}$, and  we  build the transverse action in this gauge. 

Next, we introduce the charge $Q = \int dx^N \xi^i G_i$ (recall the conditions \eqref{spatialtimeindepxi}) that generates \eqref{higherdimcoad} through $\delta_{\xi} D_{ij} = \{D_{ij} , Q \}$. The PB is evaluated with the introduced conjugate momentum $X^{ij}$ of $D_{ij}$. From this one recovers the Virasoro analog of the Gauss law operator
\al{ \label{DiffGaussLaw}
G_k = X^{ij} \partial_k D_{ij} - 2 \partial_i (X^{ij} D_{kj}) - q \partial_k \partial_i \partial_j X^{ij}. 
}
This is used to be called  the diff-Gauss law\cite{LR95,Takthesis}. The shortcut prescription to find the diff-Gauss law operator is given by
\al{ \label{diffGaussshortcutpres}
G_k = (-\delta_{\xi} D_{ij})_{\xi^k \rightarrow X^{ij0}}.
} 
Using $Q$ one can also compute 
\al{ \label{deltaXij}
\delta_{\xi} X^{ij} = \{X^{ij}, Q\}= \xi^k \partial_k X^{ij} - \partial_k \xi^i X^{kj} - \partial_k \xi^j X^{ik} + \partial_k \xi^k X^{ij}
}
The presence of the last term indicates that  $X^{ij}$ is a rank-two, spatial tensor density of weight one. Again, by its definition through the action, the momentum has a hidden time index, thereby also does the diff-Gauss law operator. Hence, from here on, we write $X^{ij0}$ and $G_k^{ \ 0}$.

We use the same prescription, \eqref{transactpresc}, with the symplectic term ST $= (\partial_0 D_{ij}) X^{ij0}$, $\mathcal{H} = X^{ij0} X_{ij0}/2$. In analogy with the YM case we choose the Lagrange multiplier as $D_0^{\ k}$. Substituting all components of the Lagrangian and performing integration by parts\footnote{As in the case of YM theory we assume the necessary boundary conditions to make total spatial derivatives vanish.} we get 
\al{ \label{diffLagmidstep}
\mathcal{L}_D = X^{ij0} ( D_{ 0}^{\ k} \partial_k D_{ij} + 2 \partial_i D_0^{\ k} D_{kj} + q \partial_i \partial_j \partial_k D_0^{\ k}) - (1/2) X^{ij0} X_{ij0}.
}

For convenience, we introduce the notation $D_{ijk} \equiv \partial_k D_{ij}, D_{ijkl} \equiv \partial_l \partial_k D_{ij}$ etc for the derivatives of the diff field only.  
In refs \cite{BLR,BRY,RY} the  diff momentum was chosen as $X^{ij0} = D^{ij0}$. However, as we show below, the action constructed from this momentum does yield back a different expression for the momentum, so  we divert from their analysis at this point.  Instead we employ the ansatz analogous to \eqref{momansKM}, namely, 
\al{ \label{Virans}
X^{ij0} = D^{ij0} + Y^{ij0}
}
with $Y^{ij0}$ being a functional of $D_{ij}, D_{0k}$ and their space derivatives, but not time derivatives. We substitute this ansatz in \eqref{diffLagmidstep} and obtain the Lagrangian in terms of $D_{ij}, D_{0k}$ and $Y_{ij0}$.  Then we recompute the momentum from it. This yields
\al{ \label{diffmomflat}
X^{ij0} \equiv \frac{\delta S_D}{\delta D_{ij0}} = D^{ij0} + D_k^{\ 0} D^{ijk} + D_k^{\ 0i} D^{k j} + D_{k}^{ \ 0 j} D^{i k} + q D_k^{\ 0 k ij}.
}
This is consistent with the ansatz \eqref{Virans}.
The shortcut prescription to find the diff momentum is 
\al{\label{shortcutmomentumprescription}
X_{ij0} = D_{ij0} + (\delta_{\xi} D_{ij})_{\xi^k \rightarrow D_0^{\ k}}.
}
With the final form of the momentum, the Lagrangian  simplifies to 
\al{ \label{diffLag}
\mathcal{L}_D  = (1/2) X^{ij0} X_{ij0}
}
which is in the form of momentum-square as the YM Lagrangian. In fact, the analogy is carried further. To see this we introduce a constant $\alpha$ for the terms involving double diff factors i.e. we write
\al{
X^{ij0} = D^{ij0} + \alpha( D_k^{\ 0} D^{ijk} + D_k^{\ 0i} D^{k j} + D_{k}^{ \ 0 j} D^{i k}) + q D_k^{\ 0 k ij}.
}
Then  we expand the Lagrangian \eqref{diffLag} and rearrange the terms   as 
\al{ \label{LagDexp}
\mathcal{L}_D = (L_1 + L_q + L_{q^2}) + (L_{\alpha} + L_{\alpha q}) + L_{\alpha^2}.
}
Here $L_1$ is the sum of the terms with no $\alpha$ nor $q$, $L_q$ with coefficient $q$, and so on. Then the first part in \eqref{LagDexp}  represents the contributions to the propagator, the second part to the three-point vertex and the last term to the four-point vertex. The structure of the actions is also the same in this respect. However, Unlike the YM case there are higher order contributions to the propagator and to the three-point vertex since the Virasoro coadjoint action is in higher order contrary to the that of KM. 

Let us also mention that the previously obtained action \cite{BRY,RY,Takthesis} corresponds to the symplectic part of $\mathcal{L}_D$. Explicitly, it is given by
\al{\label{LagBLR}
\mathcal{L}_{\text{BLR}} = (1/2) D_{ij0} \Big(D^{ij0} + \alpha( D_k^{\ 0} D^{ijk} + D_k^{\ 0i} D^{k j} + D_{k}^{ \ 0 j} D^{i k}) + q D_k^{\ 0 k ij} \Big).
}
The momentum which is used to construct this action is $X^{ij0} = D^{ij0}$. The momentum obtained from it, however, becomes 
\al{
X_{\text{BLR}}^{ij0} = D^{ij0} + (1/2) \Big(\alpha( D_k^{\ 0} D^{ijk} + D_k^{\ 0i} D^{k j} + D_{k}^{ \ 0 j} D^{i k}) + q D_k^{\ 0 k ij} \Big).
} 
This is the stated inconsistency and the basis of our modification. The Lagrangian \eqref{LagBLR} is not in the form of momentum squared and the four-point vertex is not present in it.

Recall the lie variation \eqref{deltaXij} of the diff momentum $X^{ij0}$ which showed that $X^{ij0}$ is a spatial rank-two tensor density of weight one. 
We can form a true rank-two spatial tensor out of $X^{ij0}$ as  
\al{
\tilde{X}^{ij0} = \sqrt{h} X^{ij0}
}
where $h$ is the determinant of the spatial hypersurface of the spacetime. In flat space with $h=1$ there seems to be no difference between $\tilde{X}$ and $X$. This is an illusion, however, since any variations should be done before evaluating the fields. By computing the lie variation of $\mathcal{L}_D$ it is straightforward to  show that the Lagrangian \eqref{diffLag}
is not a spatial scalar. However, the Lagrangian  
\al{ \label{spacovLagD}
\tilde{\mathcal{L}}_D = (1/2) \tilde{X}^{ij} \tilde{X}_{ij} = (1/2) h X^{ij0} X_{ij0}
}is a spatial scalar, so it 
yields a spatially invariant action
\al{ \label{SpatiallyDiffInvAction}
S = \int dt \int d^N x \, h^{3/2} \, X^{ij0} X_{ij0}. 
}
Notice that  the expression \eqref{diffmomflat} for the momentum is obtained for a flat spatial hypersurface and at this point we do not know how it may be generalized to a non-flat spatial hypersurface. In Section \ref{ModifiedTransverseAction} we provide a candidate expression. 

\subsection{Problems with Covariantization} 
\label{Covariantization}
The last step employed in Refs \cite{BRY,RY,Takthesis} was covariantizing the action, namely, lifting both  spatial and time indices to spacetime indices, and lifting the partial derivatives to covariant derivatives. 
We argue  against this step as it leads to inconsistencies.

The assumption  was that it would be possible to keep the diff-Gauss law constraint, but interpret the diff field as a tensor.  However, upon covariantization, the distinction between time and space indices is lost, and the $D_{0k}$ component, which is the Lagrange multiplier of the diff-Gauss law, becomes dynamical. Consequently the diff-Gauss law is lost as a constraint.

Suppose we  enforce the diff-Gauss law to the action at the very end. Even then there is a hidden problem: what makes the Gauss law special is not that it is a constraint, but that it is a first-class constraint generating a local transformation, i.e. residual time-independent gauge transformations. The reason the diff-Gauss law is named so, is that it is expected to be the Virasoro  analog of the Gauss law i.e. we need it to be a first-class constraint generating a local transformation.  Our expectation of the diff field theory being the gravitational analog of the YM theory  dictates this requirement. Then, as we discussed prior to equation \eqref{DiffGaussLaw} the local transformation it generates for the diff field is none other than the non-tensorial lie derivative \eqref{higherdimcoad}. In other words, the theory can not admit the diff-Gauss law as a first class constraint\footnote{Note that we do not claim that the action before covariantization has the diff-Gauss law as a first class constraint. This may require additional modifications and  will be examined in another work. We claim, on the other hand, that the diff-Gauss law is a first-class constraint if and only if the diff field transforms as in \eqref{higherdimcoad} which is not the transformation of a tensor.} and a tensorial diff field at the same time. Hence, if we treat the diff field as a tensor 
 the theory loses the connection to its origins and mimicking YM construction from KM coadjoint representation would become meaningless.  Note that these issues do not appear in the YM case. Upon covariantization,  $A_0$ remains non-dynamical, the Gauss law is still a first-class constraint generating time-independent spatial gauge transformations  and the gauge structure of the theory is unaffected.
 
 If the diff field is not a tensor, on the other hand, its covariant derivative is not well-defined. Even if we extended the definition of covariant derivative to such an object, its covariant derivative would not transform as a tensor\footnote{The covariant derivative preserves covariance, so it also preserves non-covariance.}, so contraction of indices would not yield a scalar.   This problem  also affects the interactions of the diff field introduced in \cite{BRY}. If the diff field is not a  tensor then the proposed interactions break the diffeomorphism invariance.

Below we propose a partial solution to the problem. Namely, our proposal will recover full covariance in diff-interactions, but maintain only spatial covariance in the diff field theory itself.

\section{Coadjoint Element $\Sigma$ Formed from Affine Connection }
\label{connectionVircoadjointelement}

Under a coordinate transformation $x \mapsto \bar{x}(x)$, affine connection (not necessarily Levi-Civita) coefficients $\Gamma_{ab}^{c}$ transform  as
\al{\label{gamma trf}
\bar{\Gamma}^{c}_{ab} (\bar{x})= \frac{\partial \bar{x}^{c}}{\partial x^{d}}  \frac{\partial x^{e}}{\partial \bar{x}^{a}}  \frac{\partial x^{f}}{\partial \bar{x}^{b}} \Gamma^{d}_{e f} (x) -  \frac{\partial x^{d}}{\partial \bar{x}^{a}}  \frac{\partial x^{e}}{\partial \bar{x}^{b}}  \left( \frac{\partial^2 \bar{x}^{c}}{\partial x^{d} \partial x^{e}} \right).    
}
This deviates from the transformation of a (1,2)-tensor by the last term.  Consider   
 an infinitesimal coordinate transformation $\bar{x} = x - \xi(x)$ in 1D.
We shall use the convention that if the argument of a field is suppressed, it is $x$ i.e. the original coordinate. Then, 
 \eqref{gamma trf} reduces to $
\bar{\Gamma} (\bar{x})   = \Gamma  + \Gamma \xi' +\xi''$ in 1D. 
On the other hand, by Taylor expansion we also get $\bar{\Gamma}(\bar{x}) = \bar{\Gamma} ( x- \xi)  = \bar{\Gamma}  - \xi \bar{\Gamma}' = \bar{\Gamma} - \xi \Gamma'$. Combining the two expressions we get the infinitesimal coordinate transformation of $\Gamma$ 
\al{ \label{deltagamma}
 \delta \Gamma  : = \bar{\Gamma}  - \Gamma   =  \xi \Gamma' + \Gamma \xi' +\xi''. 
}
We define the object 
\al{ 
\Sigma \equiv q(\Gamma' - \Gamma^2/2).
}
Using 
$\delta (\Gamma') = (\delta \Gamma)'$,  
$\delta (\Gamma^2/2) =  \Gamma \delta \Gamma$ and \eqref{deltagamma} it is straightforward to compute\cite{Courant} 
\al{\label{Sigmainfinitesimalcoad}
\delta \Sigma= \xi \Sigma' + 2 \xi' \Sigma + q\xi'''.
}
So $\Sigma$ 
 transforms  as a Virasoro coadjoint element with central charge $q$.
Note  that the difference of two Virasoro coadjoint elements with the same central charge transforms as a tensor. 

We would like to lift $\Sigma$ to a rank-two object  in higher dimensions.  The most general rank-two lift can be written as
\al{ \label{higherdimcoadfield}
\Sigma_{\mu \nu}  \equiv a \partial_{\lambda} \Gamma^{\lambda}_{\mu \nu} +b \partial_{\mu} \Gamma^{\lambda}_{\lambda \nu}  + c  \partial_{\nu} \Gamma^{\lambda}_{\mu \lambda} + d \Gamma^{\lambda}_{\mu \nu} \Gamma^{\sigma}_{\sigma \lambda}+ e \Gamma^{\lambda}_{\mu \sigma} \Gamma^{\sigma}_{\nu \lambda}  + f \Gamma^{\lambda}_{\mu \lambda} \Gamma^{\sigma}_{\nu \sigma} 
}
with the condition 
\al{ \label{higherdimcentralchargecond}
q \equiv a+b+c = -2 (d+e+f).
}
This leads to a higher-dimensional, rank-two lift of a Virasoro coadjoint element of central charge $q$. Recall the natural higher dimensional lift \eqref{CoadNDtrf} of Virasoro coadjoint action. 
The question is whether the object $\Sigma_{\mu \nu}$ defined in \eqref{higherdimcoadfield} transform as in \eqref{CoadNDtrf}. 
The answer turns out to be negative : 
\al{ \label{deltaLambdaansatz}
\delta \Sigma_{\mu \nu} = \xi^{\lambda} \partial_{\lambda} \Sigma_{\mu \nu} + \partial_{\mu} \xi^{\lambda} \Sigma_{\lambda \nu} + \partial_{\nu} \xi^{\lambda} \Sigma_{\mu \lambda} + q \partial_{\mu} \partial_{\nu} \partial_{\lambda} \xi^{\lambda} + \Delta_{\mu \nu}
}
where
\al{ \label{Vircoaddifference}
\Delta_{\mu \nu} & = (-a +d) \Gamma^{\rho}_{\mu \nu} \partial_{\rho} \partial_{\sigma} \xi^{\sigma} + (a+e) ( \Gamma^{\rho}_{\sigma \nu}  \partial_{\mu} \partial_{\rho} \xi^{\sigma} +  \Gamma^{\rho}_{\sigma \mu} \partial_{\rho} \partial_{\nu} \xi^{\sigma}) \notag\\ &  + f ( \Gamma^{\rho}_{\rho \nu} \partial_{\mu} \partial_{\sigma} \xi^{\sigma} + \Gamma^{\rho}_{\rho \mu} \partial_{\nu} \partial_{\sigma} \xi^{\sigma}) + (b+c+d) \Gamma^{\rho}_{\rho \sigma} \partial_{\mu} \partial_{\nu} \xi^{\sigma}. 
}
Notice that the difference only consists of terms of order $\xi''$, and the 1D reduction of $\Delta_{\mu \nu}$  subject to the condition \eqref{higherdimcentralchargecond} vanishes as expected.

\section{Spatially Covariant Extension of  Diff Field in 2D } 
\label{spatialtensor2D}

How can we recover the spatial covariance of not only the diff field action but also  the diff field itself? A quick solution that comes to mind is to complement the diff field with a correction that turns it into a tensor, just as the diff field is complemented in \eqref{gaugeinvdiff} with a correction involving the gauge field to result in a gauge invariant object. 
In fact,  the solution of this problem is simple in 1D. Based on the results of the previous section, one may subtract  $\Sigma= q(\Gamma' - \Gamma^2/2)$ from diff field $D$ (of central charge $q$) to obtain a rank-two tensor. This result, however, does not nicely extend to higher dimensions. The transformation of  $\Sigma_{\mu \nu}$   deviates from our choice \eqref{higherdimcoadfield} for building the transverse action with a rather complicated expression \eqref{Vircoaddifference}.
However, to obtain a spatial tensor from the diff field  in 2D, it is sufficient to find a $\Sigma_{\mu \nu}$ for which $\Delta_{11}$ vanishes. 
Here is one such $\Sigma_{\mu \nu}$ for $a=2, e=-1$ : 
\al{ \label{workingcase}
\Sigma_{\mu \nu}  \equiv 2 \partial_{\lambda} \Gamma^{\lambda}_{\mu \nu} - \Gamma^{\lambda}_{\mu \sigma} \Gamma^{\sigma}_{\nu \lambda} = \Sigma_{\nu \mu}. 
}
 For this choice we get a  coadjoint element of central charge two upon reduction to 1D,
and the difference from the desired variation reduces to 
\al{ \label{deltamunuofspecialsoln}
\Delta_{\mu \nu} & = -2 \Gamma^{\rho}_{\mu \nu} \partial_{\rho} \partial_{\sigma} \xi^{\sigma} +  \Gamma^{\rho}_{\sigma \nu}  \partial_{\mu} \partial_{\rho} \xi^{\sigma} +  \Gamma^{\rho}_{\sigma \mu} \partial_{\rho} \partial_{\nu} \xi^{\sigma}.
}
Following the analysis in Section \ref{TransverseMain}, the conditions $\partial_0 \xi^1 = 0 = \xi^0$ imply  vanishing of $\Delta_{11}$ automatically. Vanishing of $\Delta_{01}$ requires $
\Gamma_{01}^1 = 0$ and finally vanishing of $\Delta_{00}$ requires $\Gamma^1_{00}=0$. So we can state the analog of \eqref{spatialtimeindepxi} and  \eqref{allDscoadhigher} for \eqref{workingcase} as 
\begin{subequations}  \label{nongaugefixingSigmamunu}
\al{
\xi^0& =0 =\partial_0 \xi^1, \\
 \Gamma^1_{01} & = 0 = \Gamma^1_{00}, \\
\delta \Sigma_{11} & = \xi^1 \partial_1 \Sigma_{11} + 2 \partial_1 \xi^1 \Sigma_{11} + 2 \partial_1^3 \xi^1,  \\
\delta \Sigma_{01}  & = \xi^1 \partial_1 \Sigma_{01} + \Sigma_{01} \partial_1 \xi^1, \\
\delta \Sigma_{00}  & = \xi^1 \partial_1 \Sigma_{00}.
}
\end{subequations}

Using \eqref{nongaugefixingSigmamunu}, we can obtain a spatial tensor in 2D from the diff field by
\al{ \label{Tensoroutofdiff}
T_{\mu \nu} \equiv D_{\mu \nu} - \frac{q}{2} \Sigma_{\mu \nu}
}
It satisfies
\begin{subequations}
\al{
\xi^0& =0 =\partial_0 \xi^1, \\
&  \Gamma^1_{01} = 0 = \Gamma^1_{00}, \\
\delta T_{11} & = \xi^1 \partial_1 T_{11} + 2 \partial_1 \xi^1 T_{11}, \\  \delta T_{01} & = \xi^1  \partial_1 T_{01} + T_{01}  \partial_1 \xi^1,  \\  \delta T_{00} & = \xi^1 \partial_1 T_{00}. 
}
\end{subequations}

One may assume that this result would nicely extend to higher dimensions, but it does not. The spatial part $\Delta_{\ij}$ of \eqref{deltamunuofspecialsoln}  vanishes only in 2D.  In higher dimensions there is no obvious way to solve $\Delta_{ij}=0$. 
This  shows that although  transverse theory is spatially diffeomorphism invariant in any dimensions,  the diff field itself admits a correction to become a spatial tensor only in 2D. Extension of this result to higher dimensions requires the modification of the very first step of the construction of Section \ref{Virasorotransverse}, namely, the higher dimensional lift  \eqref{CoadNDtrf} of the coadjoint action. 
We will investigate this in the next section.

\section{Full Covariance Recovered in Interactions In Any Dimensions}
\label{DiffInvRecovered}
The analysis in the previous section suggests that the higher dimensional lift \eqref{CoadNDtrf} of the coadjoint transformation \eqref{deltaxiD}
is too strict for the diff field to be complemented with a correction turning it into a spatial  tensor. In fact,  \eqref{CoadNDtrf}  is  inconsistent as a lie derivative due to the following inequality\cite{larsson} 
\al{ \label{lieincons}
(\delta_{\eta} \delta_{\xi} - \delta_{\xi} \delta_{\eta} - \delta_{[\eta , \xi]} )D_{\mu \nu} \neq 0 
}  

Motivated with the results of the previous section, we  introduce the modified lie derivative
\al{ \label{NewVariationofDiffTensor}
\delta_{\xi} D_{\mu \nu} =  \xi^{\lambda} (\partial_{\lambda} D_{\mu \nu} )  + (\partial_{\mu} \xi^{\lambda})D_{\lambda \nu} + (\partial_{\nu} \xi^{\lambda}) D_{\mu \lambda}  +  q \partial_{\mu} \partial_{\nu} \partial_{\lambda} \xi^{\lambda}  + (q/2) \Delta_{\mu \nu}
}
with 
\al{
\Delta_{\mu \nu} & = -2 \Gamma^{\rho}_{\mu \nu} \partial_{\rho} \partial_{\lambda} \xi^{\lambda} +  \Gamma^{\rho}_{\lambda \nu}  \partial_{\mu} \partial_{\rho} \xi^{\lambda} +  \Gamma^{\rho}_{\lambda \mu} \partial_{\rho} \partial_{\nu} \xi^{\lambda}.
}
It is straightforward to show that this expression is consistent as a lie derivative i.e. equation \eqref{lieincons} turns into an equality. 
The expression \eqref{NewVariationofDiffTensor} reduces   in 1D to \eqref{deltaxiD} as desired. 
 Moreover, the object $T_{\mu \nu}$ defined in  \eqref{Tensoroutofdiff} now 
becomes a spacetime tensor in any dimension. Hence, with this new proposal we can recover full diffeomorphism invariance at least in  interactions of the diff field by using $T_{\mu \nu}$ in place of $D_{\mu \nu}$. 

We apply the same procedure as in the previous section to obtain the spatial reduction of the field components. In addition to the conditions 
\al{
\xi^0 = 0 = \partial_0 \xi^k
}
we need $\Gamma^m_{k0}=0$ to make $D_{0i}$ a spatial covariant vector and $\Gamma^m_{00}=0$ to make $D_{00}$ a spatial scalar. With these we obtain
\al{
 \xi^0 & = 0 = \partial_0 \xi^k,  
 \\
  \Gamma^m_{k0} &= 0 = \Gamma^m_{00}, 
 \\
\delta D_{ij} & = \xi^k \partial_k D_{ij} + \partial_i \xi^k D_{kj} + \partial_j \xi^k D_{ik} + q \partial_i \partial_j \partial_k \xi^k, 
\notag\\
& + (q/2) ( -2 \Gamma^m_{ij} \partial_m \partial_k \xi^k + \Gamma^m_{kj} \partial_i \partial_m \xi^k + \Gamma^m_{ki} \partial_j \partial_m \xi^k ), 
\\
  \delta D_{0i} & = \xi^k \partial_k D_{0i} + \partial_i \xi^k D_{0k}, 
\\
 \delta D_{00} & = \xi^k \partial_k D_{00}.
}
The change in $\delta D_{\mu \nu}$ will change the expression for the diff-Gauss law and as a result the expression for the  momentum. We  investigate the transverse action obtained from \eqref{NewVariationofDiffTensor} in the next section.

\section{Modified Transverse Action }
\label{ModifiedTransverseAction}
We can use the shortcut prescription \eqref{diffGaussshortcutpres} to get the modified diff-Gauss law
\al{
G^0_k & = X^{ij} \partial_k D_{ij} + \partial_i X^{ij}  D_{kj} + \partial_j X^{ij}  D_{ik} + q \partial_i \partial_j \partial_k X^{ij} \notag\\ 
& + (q/2) ( -2 \Gamma^m_{ij} \partial_m \partial_k X^{ij} + \Gamma^m_{kj} \partial_i \partial_m X^{ij} + \Gamma^m_{ki} \partial_j \partial_m X^{ij}). 
}
Since the $D$ dependent terms of the diff-Gauss law are the same, the lie derivative \eqref{deltaXij} of $X^{ij}$ is unchanged. Hence, $X^{ij}$ is still a spatial tensor density and 
\al{
\tilde{X}^{ij} = \sqrt{h} X^{ij}
}
is a rank-two spatial tensor. 
We apply the prescription \eqref{transactpresc} to find the diff-Lagrangian involving the diff momentum. We introduce the same ansatz \eqref{Virans} for the diff momentum. We insert this ansatz into the action and recompute the momentum. The result can be easily obtained from \eqref{shortcutmomentumprescription}
which yields 
\al{ \label{connectionmodifieddiffmomentum}
X_{ij0} & = \partial_0 D_{ij} +  D_0^{\ k} \partial_k D_{ij} + \partial_i D_0^{\ k} D_{kj} + \partial_j D_0^{\ k} D_{ik} + q \partial_i \partial_j \partial_k D_0^{\ k}
\notag\\
& + (q/2) ( -2 \Gamma^m_{ij} \partial_m \partial_k D_0^{\ k} + \Gamma^m_{kj} \partial_i \partial_m D_0^{\ k} + \Gamma^m_{ki} \partial_j \partial_m D_0^{\ k} ) 
}
The form of the spatially covariant Lagrangian is unchanged
\al{
\mathcal{L} = \frac{1}{2} h X_{ij0} X^{ij0}.
}

The stated similarities of the structures with YM theory also hold for the modified theory.
The action  with the modified momentum \eqref{connectionmodifieddiffmomentum} is a candidate for the spatially covariant action \eqref{SpatiallyDiffInvAction} on a spatially curved spacetime that we were looking for.
To reach  full spacetime covariance one may need to introduce objects analogous to the lapse and shift functions. It may also be required to check whether the obtained action admits the diff-Gauss law as a first-class constraint or further modifications are needed for this purpose. These will be considered in a future work. 
 
\section{Conclusions}
We showed in Section \ref{Virasorotransverse} that the diff momentum obtained from the constructed diff field action in \cite{BRY,RY,Takthesis} is not the same as the diff momentum used to construct the action. We solved this inconsistency by introducing a step, \eqref{Virans}, to the construction  procedure. With this modification the diff field action  took the form of momentum-squared \eqref{diffLag} just as in the YM action, and it had the same vertex structure as the YM theory. 

We explained in Section \ref{Covariantization} why the diff field cannot transform as a tensor if it is to be the gravitational analog of a YM field. This, however, invalidates covariantization as a way to reach a spacetime diffeomorphism invariant action. Also it invalidates the interactions of the diff field with other fields introduced in \cite{BRY}. We solved this problem by first generalizing the transformation \eqref{CoadNDtrf} of the diff field to \eqref{NewVariationofDiffTensor}, then by deriving  a tensor \eqref{Tensoroutofdiff} out of the diff field, to be used in interactions. This changed the diff field action, but it preserved the stated structural resemblance to YM action. We also showed that the modified action is still spatially diffeomorphism invariant.

%%%%%%%%%%%%%%%%%%%%%%%%%%%%%%%%%%%%%%%%
\section*{Acknowledgments}
%%%%%%%%%%%%%%%%%%%%%%%%%%%%%%%%%%%%%%%%
The author would like to thank Vincent Rodgers and Cuneyt Sahin for discussion and support.

\bibliographystyle{JHEP}
\bibliography{references}

\begin{thebibliography}{99}


\bibitem{BRY} T. Branson,  V. G. J. Rodgers, T. Yasuda, Interactions of a string inspired graviton field,
{\it Int. J. Mod. Phys. A} { 15(22)}, 3549 (2000).

\bibitem{RY} V. G. J. Rodgers, T. Yasuda, General coordinate transformations as the origins of dark energy,
{\it Int. J. Mod. Phys. A} { 22(04)}, 749 (2007). 

\bibitem{Takthesis}
T. Yasuda, The cosmological implications of diffeomorphisms, (2006)

\bibitem{WZNW} E. Witten, Non-abelian bosonization in two dimensions {\it Comm. Math. Phys.} {92(4)}, 455 (1984).


\bibitem{Pol87} A. M. Polyakov, Quantum gravity in two dimensions,
{\it Mod. Phys. Lett. A} {2(11)}, 893 (1987).

\bibitem{RR89} B. Rai and V. G. J. Rodgers, From coadjoint orbits to scale invariant wznw type actions and 2d quantum gravity action,
{\it Nucl. Phys. B} { 341(1)}, 119 (1990).

\bibitem{AS89} A. Alekseev and S. Shatashvili, Path integral quantization of the coadjoint orbits of the virasoro group and 2-d gravity,
{\it Nucl. Phys. B} { 323(3)}, 719 (1989).

\bibitem{diVecetal} P. Di Vecchia, B. Durhuus and J. L. Petersen, The Wess-Zumino action in two dimensions and non-abelian bosonization,
{\it Phys. Lett. B} { 144(3-4)}, 245 (1984).


\bibitem{Pol81} A. M. Polyakov, Quantum Geometry of Bosonic Strings,
{\it Phys. Lett. B} { 103(3)}, 207 (1981).


\bibitem{Zamo}V. G. Knizhnik, A. M. Polyakov and A. B. Zamolodchikov, Fractal structure of 2d--quantum gravity,
{\it Mod. Phys. Lett. A} { 3(08)}, 819 (1988).

\bibitem{LR95}  R. Lano and V. G. J. Rodgers, A study of fermions coupled to gauge and
gravitational fields on a cylinder,
{\it Nucl. Phys.  B} { 437(1)}, 45 (1995).

\bibitem{BLR} T. Branson, R. Lano and V. G. J. Rodgers, Yang-Mills, gravity, and string symmetries,
{\it Phys. Lett. B} { 412(3)}, 253 (1997).



\bibitem{delotez} D. Kilic, The Diffeomorphism Field {\it arXiv}\,:\,1907.06766, (2018).

\bibitem{Courant} X. Liu, L. A. P. Zayas, V. G. J. Rodgers and L. Rodriguez, A geometric action for the courant bracket
{\it arXiv:hep-th/0610021v2}   (2006).

\bibitem{larsson} T. Larsson, private communication, (Aug, 2019).





\end{thebibliography}

%%%%%%%%%%%%%%%%%%%%%%%%%%%%%%%%%%%%%%%%

\end{document}